\documentclass[12pt,a4paper]{article}
\usepackage{graphics,epsfig,rotating,amssymb}
\begin{document}
\newcommand{\goo}{\,\raisebox{-.5ex}{$\stackrel{>}{\scriptstyle\sim}$}\,}
\newcommand{\loo}{\,\raisebox{-.5ex}{$\stackrel{<}{\scriptstyle\sim}$}\,}

\begin{center}
{\Large \bf Role of multifragmentation in spallation reactions.}
\end{center}
\vspace{0.5cm}
\begin{center}
{\Large A.S.~Botvina}\\
\end{center}
\begin{center}
{\it
 Institute for Nuclear Research, Russian Academy of Sciences, 117312 Moscow,
 Russia\\
 }
\end{center}
\normalsize

\vspace{0.3cm}
\begin{abstract}
 In nuclear reactions induced by hadrons and ions of high energies,
 nuclei can disintegrate into many fragments during a short time
 ($\sim$100 fm/c). This phenomenon known as nuclear multifragmentation was
 under intensive investigation last 20 years. It was established
 that multifragmentation is an universal process taking place in all
 reactions when the excitation energy transferred to nuclei is high
 enough, more than 3 MeV per nucleon, independently on the initial
 dynamical stage of the reactions. Very known compound nucleus decay
 processes (sequential evaporation and fission), which are usual for
 low energies, disappear and multifragmentation dominates at high
 excitation energy. For this reason, calculation of multifragmentation
 must be carried on in all cases when production of highly excited
 nuclei is expected, including spallation reactions. From the other
 hand, one can consider multifragmentation as manifestation of the
 liquid-gas phase transition in finite nuclei. This gives way for
 studying nuclear matter at subnuclear densities and for applications 
 of properties of nuclear 
 matter extracted from multifragmentation reactions in astrophysics.
 In this contribution, the Statistical Multifragmentation Model (SMM),
 which combines the compound nucleus processes at low energies and
 multifragmentation at high energies, is described. The most important 
 ingredients of the model are discussed.
 \end{abstract}



 \vspace{0.5cm}

\section{Introduction}

Statistical approaches have proved to be very successful for description of 
nuclear reactions. According to the statistical hypothesis, initial dynamical 
interactions between nucleons lead to re-distribution of the available 
energy among many degrees of freedom, and the nuclear system 
evolves towards equilibrium. The most famous example of such an 
equilibrated nuclear source is the 'compound nucleus' introduced by Niels 
Bohr in 1936 \cite{Bohr}. 
It was clearly seen in low-energy nuclear reactions leading to excitation 
energies of a few tens of MeV. It is remarkable that the statistical 
concept works also for nuclear reactions induced by particles
and ions of intermediate and high energies, when nuclei break-up into many
fragments (multifragmentation) \cite{SMM}. 
In the most general consideration the process may be
subdivided into several stages: (1) a dynamical stage leading to
formation of equilibrated nuclear system, (2) disassembly of the system
into individual primary fragments, (3) de-excitation of hot primary
fragments.

\section{Formation of thermalized nuclear system}

At present, a number of dynamical models is used for description of
nuclear reactions at intermediate energies. The Intranuclear Cascade Model
was the first one used for realistic calculations of ensembles of
highly excited residual nuclei which undergo multifragmentation, see e.g.
\cite{Yariv,Botvina90,INCL4}. 
Other more sophisticated models were also used for
dynamical simulations of heavy-ion reactions, such as quantum
molecular dynamics (QMD), Boltzmann (Vlasov)-Uehling-Uhlenbeck (BUU, VUU)
and other similar models (see e.g. refs. \cite{dynmodels}).
All dynamical models agree that the character of the dynamical
evolution changes after a few rescatterings of incident nucleons, when high
energy particles ('participants') leave the system.
This can be seen from distributions of nucleon velocities and density
profiles in remaining spectators \cite{buusmm,Bondorf94,larionov}.
However, the time needed for equilibration and transition to the
statistical description is still under debate. This time is estimated
around or less than 100 fm/c for spectator matter, however, it slightly
varies in different models. Parameters of the
predicted equilibrated sources, i.e. their excitation energies, mass
numbers and charges vary significantly with this time.
In this case a reasonable strategy is to use results of the dynamical
simulations as a qualitative guide line, but extract parameters of
thermalized sources from the analysis of experimental data.
In this case, one can avoid uncertainties of dynamical models in 
description of the thermalization process.

\section{Break-up of nuclear system into hot primary fragments}

\subsection{Evolution from sequential decay to simultaneous break-up}

After dynamical formation of a thermalized source, its further evolution
depends crucially on the excitation energy and mass number.
The standard compound nucleus picture is valid only at low excitation
energies when sequential evaporation of light particles and fission are
the dominant decay channels \cite{SMM}. 
Some modifications of the evaporation/fission
approach were proposed in order to include emission of fragments heavier
than $\alpha$-particles, see e.g. \cite{Moretto,Botvina87,charity0}.
However, the concept of the compound nucleus cannot be applied at high
excitation energies, $E^* \goo$ 3 MeV/nucleon. The reason is that the time
intervals between subsequent fragment emissions, estimated both within the
evaporation models \cite{charity} and from experimental data \cite{jandel},
become very short, of order of a few tens of fm/c. In this case there will
be not enough time for the residual nucleus to reach equilibrium between
subsequent emissions.
Moreover, the produced fragments will be in the vicinity of each other and,
therefore, should interact strongly. The rates of the particle emission
calculated as for an isolated compound nucleus will not be reliable
in this situation. There are many other theoretical arguments 
in favour of a simultaneous break-up at high excitation energy. For 
example, the Hartree-Fock and Thomas-Fermi calculations predict that 
the compound nucleus will be unstable at high temperatures \cite{HFTF}. 
Sophisticated dynamical calculations have also shown that a nearly
simultaneous break-up into many fragments is the only possible way for
the evolution of highly-excited systems \cite{XXX}. 
 
On the other hand, the picture of a nearly simultaneous
break-up in some freeze-out volume is more justified in this case.
Indeed, the time scales of less than 100 fm/c are extracted for
multifragmentation reactions from experimental data
\cite{beaulieu,karnaukhov}.
There exist several analyses of experimental data, which also reject the
binary decay mechanism of fragment production
via sequential evaporation from a compound nucleus at high excitation energy.
For example, this follows from the fact that a popular
sequential GEMINI code cannot describe the multifragmentation data
\cite{hubele,Deses,napolit}. We believe that a formal reason of this
failure is that the evaporation approaches always predict larger
probabilities for emission of light particles (in particular, neutrons)
than for intermediate mass fragments (IMFs). The GEMINI model, which assumes
independent evaporation of fragments, fails also to describe angular
correlations of the produced IMFs \cite{ISIS}. The reason is that at 
multifragmentation these correlations reflect Coulomb interaction of many 
fragments, but not a two-body kinematic.

\subsection{Statistical multifragmentation model} 

Several versions of the statistical approach
have been proposed for the description of multifragmentation reactions
(see e.g. \cite{SMM,Randrup,Gross}). 
As the main de-excitation code we take the Statistical Multifragmentation 
Model (SMM), fully described in a review \cite{SMM}. The reason is that this 
model was primary constructed for using after initial dynamical stage, and 
adjusted for this kind of hybrid calculations. 

The model assumes statistical equilibrium of excited nuclear system with 
mass number $A_0$, charge $Z_0$, and excitation energy (above the ground 
state) $E_0$ at a low-density freeze-out volume. This volume can be 
parameterized as $V=V_0+V_f$, so the baryon density is 
$\rho = A_0 / V$. $V_0$ is the volume of the system at the normal nuclear 
density $\rho_0 \approx$ 0.15 fm$^{-3}$. $V_f$ is the so-called free 
volume available for translational motion of fragments. In the excluded 
volume approximation $V_f$ may be taken as a constant for all break-up 
channels, however, under more realistic assumption, it 
depends on fragment multiplicity $M$ in the channels \cite{SMM}. 

The model considers all break-up channels (ensemble of partitions $\{p\}$) 
composed of nucleons and excited fragments taking into account the
conservation of baryon number, electric charge and energy. 
An important advantage of the SMM is that besides these break-up 
channels it includes also the compound nucleus channel, and takes into 
account competition between all channels. In this way the SMM includes 
the conventional evaporation and fission processes at low excitation 
energy, and provides natural generalization of the de-excitation process 
for high excitation energy. 

In the model light nuclei with mass number $A \leq 4$ and charge 
$Z \leq 2$ are treated as elementary stable particles with masses and 
spins taken from the nuclear tables ("nuclear gas"). Only translational 
degrees of freedom of these particles contribute to the entropy of the 
system. Fragments  with $A > 4$ are treated as heated nuclear liquid 
drops. In this way one may study the nuclear liquid-gas coexistence in 
the freeze-out volume. Their 
individual free energies $F_{AZ}$ are parameterized as a sum of the bulk, 
surface, Coulomb and symmetry energy contributions 
\begin{equation}
F_{AZ}=F^{B}_{AZ}+F^{S}_{AZ}+E^{C}_{AZ}+E^{sym}_{AZ}.
\end{equation}
The standard expressions for these terms are:
$F^{B}_{AZ}=(-W_0-T^2/\epsilon_0)A$, where $T$ is the temperature,
the parameter $\epsilon_0$ is related to the level density, and
$W_0 = 16$~MeV is the binding energy of infinite nuclear matter;
$F^{S}_{AZ}=B_0A^{2/3}(\frac{T^2_c-T^2}{T^2_c+T^2})^{5/4}$, where
$B_0=18$~MeV is the surface coefficient, and $T_c=18$~MeV is the critical
temperature of infinite nuclear matter; $E^{C}_{AZ}=cZ^2/A^{1/3}$, where
$c=(3/5)(e^2/r_0)(1-(\rho/\rho_0)^{1/3})$ is the Coulomb parameter (obtained 
in the Wigner-Seitz approximation), with the charge unit $e$ and 
$r_0$=1.17 fm;  $E^{sym}_{AZ}=\gamma (A-2Z)^2/A$, where
$\gamma = 25$~MeV is the symmetry energy parameter.
These parameters are those of the Bethe-Weizs\"acker formula and correspond
to the assumption of isolated fragments with normal density in the
freeze-out configuration, an assumption found to be quite successful in
many applications. It is to be expected, however, that in a
more realistic treatment primary fragments will have to be considered
not only excited but also expanded and still subject to a residual nuclear
interaction between them.
These effects can be accounted for in the fragment
free energies by changing the corresponding liquid-drop parameters. 
The Coulomb interaction of fragments in the freeze-out volume is
described within the Wigner-Seitz approximation (see ref. \cite{SMM} 
for details). 

As is well known, the number of partitions
of medium and heavy systems $(A_0\sim 100)$ is enormous
(see e.g. \cite{Jackson}). In order to take them into account the 
model uses few prescriptions. 
At small excitation energies the standard SMM code \cite{SMM} uses a
microcanonical treatment, however, taking into account a limited
number of disintegration channels: as a rule, only partitions with total
fragment multiplicity
$M \leq 3$ are considered. This is a very reasonable approximation at
low temperature, when the compound nucleus and low-multiplicity channels
dominate. Recently, a full microcanonical version of the SMM using
the Markov Chain method was introduced \cite{Jackson,Botvina01}. It can
be used for exploring all partitions without limitation.
However, it is a more time consuming approach, and it is used in special
cases only \cite{Botvina01}. 

Within the microcanonical ensemble the
statistical weight of a partition $p$ is calculated as
\begin{eqnarray}
W_{\rm p} \propto exp~S_{\rm p},
\end{eqnarray}
where $S_{\rm p}$ is the corresponding entropy, which depends on fragments 
in this partition, as well as on the excitation energy $E_0$, mass 
number $A_{0}$, charge $Z_{0}$, volume $V$ of the system. In the standard 
treatment we follow a description which corresponds to approximate 
microcanonical ensemble. Namely, we introduce a temperature $T_{p}$ 
characterising all final states in each partition $p$. It is determined 
from the energy balance equation taking into account the total excitation 
energy $E_0$ \cite{SMM}. In the following we determine $S_{\rm p}$ for the 
found $T_{p}$ by using conventional thermodynamical relations. In 
the standard case, it can be written as 
\begin{eqnarray}
S_{\rm p}=ln(\prod_{A,Z}g_{A,Z})+ln(\prod_{A,Z}A^{3/2})
-ln(A_0^{3/2})-ln(\prod_{A,Z}n_{A,Z}!)+
\nonumber
\\
(M-1)ln(V_f/\lambda_{T_{p}}^3)
+1.5(M-1)+\sum_{A,Z}(\frac{2T_{p}A}{\epsilon_0}-
\frac{\partial F^{S}_{AZ}(T_{p})}{\partial T_{p}}) , \nonumber 
\end{eqnarray}
where $n_{A,Z}$ as the number of fragments with mass $A$ and charge 
$Z$ in the partition, 
$g_{A,Z}=(2s_{A,Z}+1)$ is the spin degeneracy factor, 
$\lambda_{T_{p}}=\left(2\pi\hbar^2/m_NT_{p}\right)^{1/2}$ is the nucleon 
thermal wavelength ($m_N\approx 939$ MeV is the average nucleon mass), 
and the summation is performed over all fragments of the partition $p$.
We enumerate all considered partitions and select one of them according
to its statistical weight by the Monte-Carlo method.

At high excitation energy the standard SMM code makes a transition to
the grand-canonical ensemble \cite{SMM}, since the number of partitions 
with high probability becomes too large. 
In the grand canonical formulation, after integrating out translational 
degrees of freedom, one can write the mean multiplicity of nuclear 
fragments with $A$ and $Z$ as
\begin{eqnarray}
\label{naz} \langle n_{A,Z} \rangle =
g_{A,Z}\frac{V_{f}}{\lambda_T^3}A^{3/2} {\rm
exp}\left[-\frac{1}{T}\left(F_{AZ}(T,V)-\mu A-\nu Z\right)\right]. 
\end{eqnarray}
Here the temperature $T$ can be found from the total energy balance of the 
system by taking into account all possible fragments with $A$ from 1 
to $A_0$ and with $Z$ from 0 to $Z_0$ \cite{SMM}. 
The chemical potentials $\mu$
and $\nu$ are found from the mass and charge constraints:
\begin{equation} \label{eq:ma2}
\sum_{A,Z}\langle n_{A,Z}\rangle A=A_{0},~~ 
\sum_{A,Z}\langle n_{A,Z}\rangle Z=Z_{0}. \nonumber
\end{equation}
In this case the grand canonical occupations $\langle n_{A,Z} \rangle$ are 
used for Monte-Carlo sampling of the fragment partitions \cite{SMM}. 
These two methods of partition generation are carefully adjusted to provide 
a smooth transition from the low energy to the high energy regimes.

\section{Propagation and de-excitation of hot fragments} 

After the Monte-Carlo generation of a partition the temperature of the hot 
fragments, their excitation energy and momenta can be found from the energy 
balance. In this approach the temperature may slightly fluctuate from 
partition to partition, since the total energy energy of the system 
$E_0$ is always conserved. At the next stage 
Coulomb acceleration and propagation of fragments must be 
taken into account. For this purpose the fragments are placed randomly 
in the freeze-out volume $V$ (without overlapping), and their positions are 
adjusted by taking into 
account that their Coulomb interaction energy must be equal to the value 
calculated in the Wigner-Seitz approximation. We note that in the case 
of the Markov Chain SMM version \cite{Botvina01} this adjustment is not 
necessary, since positions of fragments are sampled directly. 
In the freeze-out volume a possible collective flow of 
fragments can be also taken into account \cite{SMM}. Usually it is 
done by adding additional radial velocities to the fragments (proportional 
to their distances from the centre of mass) in the beginning of Coulomb 
acceleration. In the following we resolve the Hamilton equations for motion 
of fragment from these initial positions in their mutual Coulomb field. 
The energy and momentum balances are strictly respected 
during this dynamical propagation. 

The secondary de-excitation of primary hot fragments includes 
several mechanisms. For light primary  fragments  (with  $A\leq 16$) 
produced in multifragmentation 
even a  relatively  small  excitation  energy  may  be  comparable 
with their total binding energy. In this case we  assume  that  the 
principal mechanism of de-excitation is the explosive decay of the 
excited nucleus into several smaller  clusters  (the  Fermi 
break-up) \cite{Botvina87,SMM}.
In this decay the statistical weight of the channel $p$ containing
$n$  particles  with  masses $m_{i}$ ($i=1,\cdots,n$) in volume
$V_{p}$ can be calculated  in microcanonical approximation : 
\begin{equation} \label{eq:Fer}
\Delta\Gamma_{p}\propto
\frac{S}{G}\left(\frac{V_{p}}{(2\pi\hbar)^{3}}\right)^{n-1}
\left(\frac{\prod_{i=1}^{n}m_{i}}{m_{0}}\right)^{3/2}\frac{(2\pi)^
{\frac{3}{2}(n-1)}}{\Gamma(\frac{3}{2}(n-1))}\cdot
\left(E_{kin}-U_{p}^{C}\right)^{\frac{3}{2}n-\frac{5}{2}},
\end{equation}
where $m_{0}=\sum_{i=1}^{n}m_{i}$ is the mass of the decaying nucleus,
$S=\prod_{i=1}^{n}(2s_{i}+1)$ is the degeneracy factor 
($s_{i}$ is the $i$-th particle spin), $G=\prod_{j=1}^{k}n_{j}!$ 
is the particle identity factor ($n_{j}$ is the number of particles of 
kind $j$). $E_{kin}$ is the total kinetic energy of  particles at 
infinity which can be found through the energy balance by taking 
into account the fragment excitation energy, $U_{p}^{C}$ is the Coulomb 
barrier for this decay. We have slightly modified this model \cite{Botvina87} 
by including fragment excited states stable with respect to the 
nucleon emission as well as some long-lived unstable nuclei. 
 
The successive particle emission from hot primary fragments with 
$A>16$ is assumed to be their basic de-excitation mechanism, as in the 
case of the compound nucleus decay. 
Due to the high excitation energy of these fragments, the standard 
Weisskopf evaporation scheme was modified to take into account the 
heavier ejectiles up to $^{18}$O, besides light particles (nucleons, 
$d$, $t$, $\alpha$), in ground and particle-stable excited states 
\cite{Botvina87}. The width for the emission of a particle $j$ from 
the  compound  nucleus $(A,Z)$ is given by:
\begin{equation} \label{eq:eva}
\Gamma_{j}=\sum_{i=1}^{n}\int_{0}^{E_{AZ}^{*}-B_{j}-\epsilon_{j}^{(i)}}
\frac{\mu_{j}g_{j}^{(i)}}{\pi^{2}\hbar^{3}}\sigma_{j}(E)
\frac{\rho_{A^{'}Z^{'}}(E_{AZ}^{*}-B_{j}-E)}{\rho_{AZ}(E_{AZ}^{*})}EdE.
\end{equation}
Here the sum is taken over the ground and all particle-stable excited states 
$\epsilon_{j}^{(i)}~(i=0,1,\cdots,n)$ of the fragment $j$, 
$g_{j}^{(i)}=(2s_{j}^{(i)}+1)$   is  the 
spin degeneracy factor of the $i$-th excited  state, 
$\mu_{j}$ and $B_{j}$ are corresponding reduced mass and separation energy, 
$E_{AZ}^{*}$ is the excitation energy of the initial nucleus, 
$E$ is the kinetic energy of an emitted particle in the centre-of-mass 
frame. In eq. (\ref{eq:eva}) $\rho_{AZ}$ and $\rho_{A^{'}Z^{'}}$ are 
the level densities of the initial $(A,Z)$ and final $(A^{'},Z^{'})$ 
compound  nuclei.  The cross section $\sigma_{j}(E)$ of the inverse 
reaction $(A^{'},Z^{'})+j=(A,Z)$ was calculated using the optical model 
with nucleus-nucleus potential \cite{Botvina87}. The evaporation 
process was simulated by the Monte Carlo method and the conservation of 
energy and momentum was strictly controlled in each emission step. 

An important channel of de-excitation of heavy nuclei ($A>100$) is
fission. This process competes with particle emission, and it is also 
simulated with the Monte-Carlo method. 
Following the Bohr-Wheeler statistical approach we assume that
the partial width for the compound nucleus fission is proportional
to the level density at the saddle point $\rho_{sp}(E)$ \cite{SMM} :
\begin{equation} \label{eq:fis}
\Gamma_{f}=
\frac{1}{2\pi\rho_{AZ}(E_{AZ}^{*})}\int_{0}^{E_{AZ}^{*}-B_{f}}
\rho_{sp}(E_{AZ}^{*}-B_{f}-E)dE,
\end{equation}
where $B_{f}$ is the height of the fission barrier which is determined by
the Myers-Swiatecki prescription. For approximation of
$\rho_{sp}$ we used the results of the extensive analysis of nuclear
fissility and $\Gamma_{n}$/$\Gamma_{f}$ branching ratios, see ref. 
\cite{SMM} for details. 

All these models for secondary de-excitation 
were tested by numerical comparisons with experimental data on 
decay of compound nuclei with excitation energies less than 2--3 MeV 
per nucleons. It is important that after all stages the SMM provides 
event by event simulation of the whole break-up process and allows for 
direct comparison with experimental events.

\section{Verification and applications of the SMM}

As was shown already in first publications \cite{SMM,Botvina90} 
the SMM gives very good description
of experimental data in the case when fragments are emitted from
equilibrated sources. Later on, many experimental groups have 
successfully applied SMM for interpretation of their data: ALADIN (GSI,
Germany) \cite{ALADIN}, EOS (Purdue University, USA) \cite{EOS},
ISIS (Indiana University, USA) \cite{ISIS}, Miniball-Multics (MSU,
USA and INFN, Italy) \cite{MSU}, INDRA (GANIL, France)
\cite{INDRA}, FAZA (Dubna, Russia) \cite{FAZA}, and others.
In particular, the SMM describes charge (mass) distributions of 
produced fragments and their evolution with excitation energy, 
isotope distributions, multiplicities of produced particles and 
fragments, charge distributions of first, second, third fragments in 
the system, correlation functions (charge, angle, velocity ones) of 
the fragments, fragment kinetic energy distributions. Simultaneously, 
this model reproduces global characteristics of the systems, such as 
caloric curves, critical indexes for the phase transition, different 
moments of the fragment charge distribution. Otherwords, the model can 
describe nearly completely experimental events, and, in some cases, it can 
be used as event generator. 
Importance of multifragmentation channels for nuclear reactions is
now widely recognized. The SMM is included in many
complex codes designed to describe transport of particles and
isotope production in matter, for example, in GEANT4 (CERN)
\cite{GEANT4}.

From the other hand, systematic studies of multifragmentation 
have brought important information about 
the nuclear liquid-gas phase transition \cite{Pochodzala,Dagostino2}.
Multifragmentation reaction opens the unique possibility for investigating 
the phase diagram of nuclear matter at temperatures $T \approx 3-8$ MeV
and densities around $\rho \approx 0.1-0.3 \rho_0$, which are expected in the
freeze-out volume. Previously, only theoretical calculations without 
experimental verification were available for this phase diagram region. 
This information is crucial, for example, for construction of a reliable 
equation of state of stellar matter and modelling nuclear composition in 
supernovae explosions, where the same thermodynamical conditions of nuclear 
matter exist \cite{Botvina04}.

One of the promising application of multifragmentation reactions is 
investigation of properties of excited nuclei embedded in surrounding of 
other nuclear species. They can be modified in comparison with the properties 
of isolated nuclei, since in the freeze-out volume fragments can interact with 
each other with Coulomb and residual nuclear forces. This study can not be 
performed in conventional nuclear reactions involving only a compound 
nucleus channel. However, the new properties of fragments can be extracted 
from analysis of experimental multifragmentation data. As was found there are 
essential modifications of symmetry and surface energies of hot fragments 
\cite{LeFevre,Iglio,Botvina06,Souliotis}. These modified properties of 
fragments should be also taken into account during their secondary 
de-excitation \cite{nihal}.

\section{Conclusion} 

Nearly 60 years ago nuclear physicers started to investigate the spallation 
reaction. Primary it was considered as emission of few nucleons and light 
charged particles (and, eventually, fission) from a heavy nucleus. Only 
these processes were clearly observed that time, since the accelerators 
could provide projectile beams with relatively low energy (few hundred 
MeV). During last 15 years we have obtained solid evidences that at 
high projectile energies, and in heavy-ion collisions, a heavy nucleus 
can be completely disintegrated into light and intermediate mass 
fragments. This multifragmentation reaction is universal, and it is a 
natural fast decay taking place at high energies. The multifragmentation 
channels take as much as 10--15\% of the total cross section in high-energy 
hadron-nucleus reactions, and about twice more in high-energy nucleus-nucleus 
collisions. Moreover, multifragmentation reactions are responsible for 
production of most intermediate mass fragments and some specific isotopes. 

The traditional evaporation and fission models can not describe correctly 
this fast multifragmentation, since they are based on the 
hypothesis of a long-lived compound nucleus. There is a statistical 
approach, realized in the Statistical Multifragmentation Model (SMM), 
which allows for natural extension of conventional cascade-evaporation 
calculations for the multifragmentation reaction. At low excitation 
energy it includes compound nucleus processes, however, at high 
excitations it describes the simultaneous break-up. Already at 
present the SMM demonstrates very good description of 
experimental data, especially at high excitation energy of nuclear 
systems (more than 
3 MeV/nucleon). The problems, which are necessary to resolve within 
this approach, concern mainly a better description of transition 
from the compound nucleus to the multifragmentation decay. This is 
important for calculation of reactions initiated by low energy projectiles 
(up to 1 GeV), when very few equilibrated nuclei have a high excitation 
energy sufficient for multifragmentation.


\end{document}